# From Data Harvesting to Querying for Making Urban Territories Smart

Genoveva Vargas-Solar[1,5], Ana-Sagrario Castillo-Camporro[2,5]

José Luis Zechinelli-Martini[3,5], Javier A. Espinosa-Oviedo[4,5]

[1]*University Grenoble Alpes, CNRS, Grenoble INP, LIG, France*
[2]*Universidad Nacional Autónoma de México, Mexico*
[3]*Fundación Universidad de las Américas Puebla, Mexico*
[4]*University of Lyon, LIRIS, France*
[5]*French Mexican Laboratory of Informatics and Automatic Control*

{genoveva.vargas, javier.espinosa}@imag.fr
sagrariocastillo@comunidad.unam.mx
joseluis.zechinelli@udlap.mx

This chapter provides a summarized, critical and analytical point of view of the data-centric solutions that are currently applied for addressing urban problems in cities. These solutions lead to the use of urban computing techniques to address their daily life issues. Data-centric solutions have become popular due to the emergence of data science. The chapter describes and discusses the type of urban challenges and how data science in urban computing can face them. Current solutions address a spectrum that goes from data harvesting techniques to decision making support. Finally, the chapter also puts in perspective families of strategies developed in the state of the art for addressing urban problems and exhibits guidelines that can lead to a methodological understanding of these strategies.

# 1. Introduction

The development of digital technologies in the different disciplines, in which cities operate, either directly or indirectly, is altering expectations among those in charge of the local administration. Every city is a complex ecosystem with subsystems to make it work such as work, food, clothes, residence, offices, entertainment, transport, water, energy etc. With the growth of cities, there is more chaos and most decisions are politicized, there are no common standards and data is overwhelming. The intelligence is sometimes digital, often analogue, and almost inevitably human.

Urban computing [36] is a world initiative leading to better exploit resources in a city to offer higher-level services to people. It is related to sensing the city's status and acting in new intelligent ways at different levels: people, government, cars, transport, communications, energy, buildings, neighbourhoods, resource storage, etc. A vision of the city of the "future", or even the city of the present, rests on the integration of science and technology through information systems.

Data-centric solutions are in the core of urban computing that aims at understanding events and phenomena emerging in urban territories, predict their behaviour and then use these insights and foresight to make decisions. Data analytics and exploitation techniques are applied in different conditions and using ad hoc methodologies using data collections of different types. Today important urban computing centres in metropolises, have proposed and applied these techniques in these cities for studying real state, tourism, transport, energy, air, happiness, security and wellbeing. The adopted strategies have to do with the type of context in which they work.

This chapter provides a summarized, critical and analytical point of view of the data-centric solutions that are currently applied for addressing urban problems in cities leading the use of urban computing techniques to address their daily life issues. The chapter puts in perspective families of strategies developed in the state of the art for addressing given urban problems and exhibits guidelines that can lead to a methodological understanding of these strategies. Current solutions address a spectrum that goes from data harvesting techniques to decision making support. The chapter describes them and discusses their main characteristics.

Accordingly, the chapter is organised as follows. Section 2 characterises urban data and introduces data harvesting techniques used for collecting urban data. Section 3 discusses approaches and strategies for indexing urban data. Section 4 describes urban data querying. Section 5 summarizes data and knowledge fusion techniques. Finally, Section 6 discusses the research and applied perspectives of urban computing for concluding the chapter.

## 2. Data Harvesting Techniques in Urban Computing

Urban computing is an interdisciplinary field which concerns the study and application of computing technology in urban areas. A new research opportunity emerges in the database domain for providing methodologies, algorithms and systems to support data processing and analytics processes for dealing with urban computing. These processes involve harvesting data about the urban environment to help improve the quality of life for people in urban territories like cities. In this context academic and industrial contributions have proposed solutions for building networks of data, retrieving, analysing and visualizing them for fulfilling analytics requirements stemming from urban computing studies and projects.

Urban data processing is done using: (i) continuously harvested observations of the geographical position of individuals (that accept sharing their position) along time; (ii) collections of images stemming from cameras observing specific "critical" urban areas, like terminals, airports, public places and government offices; (iii) data produced by social networks and applications like Twitter, Facebook, Waze and similar. Independently of the harvesting strategies and processing purposes, it is important to first characterise urban data. This is done in the next section.

*2.1 Urban Data*

Urban data can be characterized concerning three properties: time, space and objects (occupying urban territories). They are elementary properties that can guide the way urban data can be harvested and then processed for understanding urban phenomena. For urban data, time must be considered from two perspectives, as its mathematical definition as a continuous or discrete linearly ordered set consisting of time instants or time intervals, called *time units* [3]. But, also under a cyclic perspective to consider iterations of seasons, weeks and days. Regarding *space*, it can be represented by [3] different referencing models: coordinate-based models with tuples of numbers representing the distance to certain reference points or axes; division-based models using a geometric or semantic-based division of space; and linear models with relative positions along with linear reference elements, such as streets, rivers and trajectories. Finally, the third urban data property, object, refers to physical and abstract entities having a certain position in space (e.g., vehicles, persons and facilities); temporal properties, for objects existing in a certain period (i.e., event); spatio-temporal properties, which are objects with a specific position in both space and time.

Besides time, space and object properties, Yixian Zheng et al. [25] identify six types of data that can be harvested and represent the type of entities that can be observed within urban territories according to the urban context they refer to human mobility, social network, geographical, environmental, health care and divers.

*Human mobility data* enables the study of social and community dynamics based on different data sources like traffic, commuting media, mobile devices and geotagged social media data.

Traffic data produced by sensors installed in vehicles or specific spots around the city (e.g., loop sensors, cameras). These data can include vehicles' positions observed recurrently at given intervals. Using these points (positions), it is then possible to compute trajectories which are spatiotemporally time-stamped and can be associated with instant speed and heading directions. Traffic occupation inroads can be measured with loops that compute within given time intervals, which vehicles travel across two consecutive loops. Using this information, it is possible to compute travel speed and traffic volume on roads. Ground truth traffic conditions are observed using surveillance cameras, that generate a huge volume of images and videos. Extracting information such as traffic volume and flowrate from these images and videos is still challenging. Therefore, in general, these data only provide a way to monitor citywide traffic conditions manually.

People's regular movement data are produced by personalized RFID transportation cards for buses or metro that they tap in station entries to enter/exit the public transportation system. This generates a huge amount of records of passenger trips where each record includes an anonymous card ID, tap-in/out stops, time, fares for this trip and transportation type (i.e. bus or metro). Commuting data recording people's regular movement in cities can be used to improve public transportation and to analyse citywide human mobility patterns.

Records of exchanges like phone calls, messages, internet, between mobile phones and cell stations collected by telecom operators are data that contain communication information, people's locations based on cell stations. These data offer unprecedented information to study human mobility.

*Social Networks Data*. Social networks posts (e.g. blogs, tweets) are tagged with geo-information that can help to better understand people's activities, the relations among people and the social structure of specific communities. User-generated texts, photos and videos, contain rich information about people's interests and characteristics, that can be studied from a social perspective. For example, evolving public attention on topics and spreading of anomalous information. The major challenges with geo-tagged social network data lie in their sparsity and uncertainty.

Finally, data refer to points of interest (POI) to depict information of facilities, such as restaurants, shopping malls, parks, airports, schools and hospitals in urban spaces. Each facility is usually described by a name, address, category and a set of geographical coordinates.

*Environmental data.* Modern urbanization based on technology has led to environmental problems related to energy consumption and pollution. Data can be produced by monitoring systems observing environment through different variables and observations (e.g. temperature, humidity, sunshine duration and weather conditions), air pollution data, water quality data and satellite remote sensing data, electricity and energy consumption, $CO_2$ footprints, gas. These data can help to have insight on consumption patterns, on correlations among actions and implications and foresight about the environment.

*Divers data*. Other data are complementary to urban data, particularly those concerning social and human aspects, such as health care, public utility service, economy, education, manufacturing and sports.

Figure 1 summarizes the urban data types considered in urban computing: environmental monitoring data that concern meteorological data; mobile phone signals used for identifying behaviours, citywide human mobility and commuting data for detecting urban anomalies, city's functional regions and urban planning; geographical data concerning points of interest (POI), land use; traffic data; social networks data; energy data obtains from sensors; and economies regarding city economic dynamics like transaction records of credit cards, stock prices, housing prices and people's income.

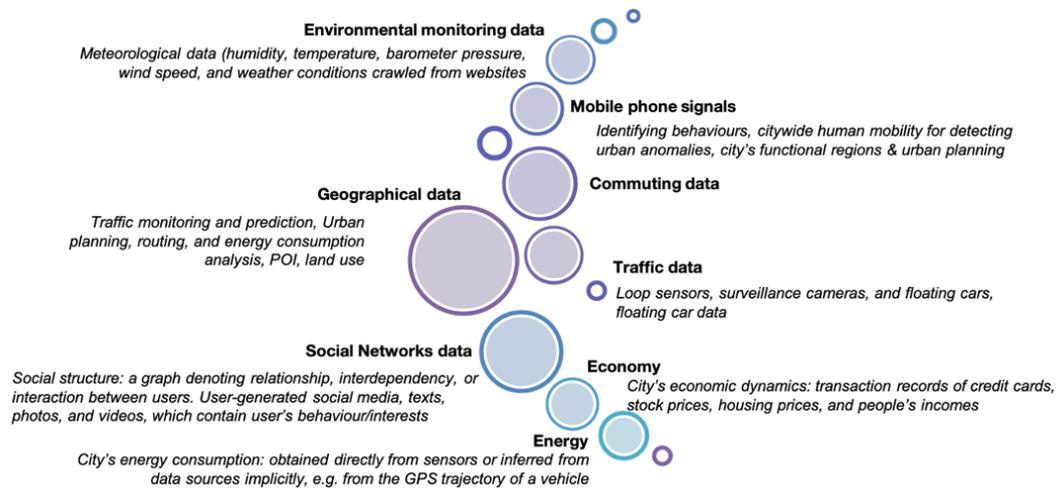

*Figure 1 Urban Data Types*

Urban data can be harvested from different sources and using different techniques. These aspects are discussed next.

## 2.2 Data Harvesting Techniques

Data acquisition techniques can *unobtrusively* and *continually* collect data on a *citywide* scale. Data harvesting is a nontrivial problem given the three aspects to consider: (i) e*nergy consumption and privacy, (ii) loose-controlled and nonuniform distributed sensors, (iii) unstructured, implicit, and noise data.*

*Crowdsensing.* The term "crowdsourcing" is defined as the practice of obtaining needed services or content by soliciting contributions from a large group of people. People play the role of urban data consumers, but also participate in the data analysis process through crowdsourcing. Techniques use explicit and implicit crowdsourcing for collecting data that contain information about the way people evolve in public and private places. These data collections can be used as input for learning crowd behaviour and simulating it more accurately and realistically.

The advances of location-acquisition technologies like GPS and Wi-Fi has enabled people to record their location history with a sequence of time-stamped locations, called trajectories. Regarding non-obstructive data harvesting, work has been carried out using cellular networks for user tracking, profiting from call delivery that uses transitions between wireless cells. Geolife[1] is a social networking service which aims to understand trajectories, locations and users, and mine the correlation between users and locations in terms of user-generated GPS trajectories. In [17] a new vision has been proposed regarding the smart cities' movement, under the hypothesis that there is the need to study how people psychologically perceive the urban environment, and to capture that quantitatively. Happy Maps uses crowdsourcing and geo-tagged pictures and the associated metadata to build alternative cartography of a city weighted for human emotions. People are more likely to take pictures of historical buildings, distinctive spots and pleasant streets instead of car-infested main roads. On top of that, Happy Maps adopts a routing algorithm that suggests a path between two locations that is the shortest route that maximizes the emotional gain.

*2.3 Discussion and Synthesis*

An important aspect to consider is that data is non-uniformly distributed in geographical and temporal spaces, and it is not always harvested homogeneously according to the technique and also the conditions of the observed entities in an urban territory.

---

[1] https://www.geospatialworld.net/article/geo-life-health-smart-city-gis/

Having the entire dataset may be always infeasible in an urban computing system. Some information is transferrable from the partial data to the entire dataset, for example, the travel speed of taxis on roads can be transferred to other vehicles that are also travelling on the same road segment. Some information cannot be transferable, for example, the traffic volume of taxis on a road may be different from private vehicles

In some locations, when crowdsensing is used, more data can be harvested as required and in other places fewer data than required. In the first case, a down-sampling method, e.g., compressive sensing, could be useful to reduce a system's communication loads. In the last case, in the context of crowdsensing, some incentives can motivate users to contribute with data should be considered. How to configure the incentive for different locations and periods to maximize the quality of the received data (e.g., the coverage or accuracy) for a specific application is yet to explore.

Three types of strategies can be adopted for harvesting data. (i) Traditional sensing and measurement that implies installing sensors dedicated to some applications. (ii) Passive crowdsensing using wireless cellular networks built for mobile communication between individuals to sense city dynamics (e.g., predict traffic conditions and improve urban planning). We described how this technique can be specialised into three strategies:

- Sensing City Dynamics with GPS-Equipped Vehicles: mobile sensors continually probing the traffic flow on road surfaces processed by infrastructures that produce data representing city-wide human mobility patterns.
- *Ticketing Systems of Public Transportation* (e.g., model the city-wide human mobility using transaction records of RFID-based cards swiping).
- *Wireless Communication Systems* (e.g., call detailed records CDR).

- *Social Networking Services (*e.g., geotagged posts/photos, posts on natural disasters analysed for detecting anomalous events and mobility patterns in the city)

(iii) Participatory sensing where people obtain information around them and contribute to formulating collective knowledge to solve a problem (i.e., human as a sensor):

- *Human crowdsensing*: users willingly sense information gathered from sensors embedded in their own devices (e.g., GPS data from a user's mobile phone used to estimate real-time bus arrivals).
- Human crowdsourcing: users are proactively engaged in the act of generating data: reports on accidents, police traps, or any other road hazard (e.g. Waze), citizens turning into cartographers, to create open maps of their cities.

## 3. Managing and Indexing Urban Data

The objective of managing and indexing urban data is to harness a variety of heterogeneous data to quickly answer users' instant queries, e.g. predicting traffic conditions and forecasting air pollution. Three problems are addressed in this context: stream and trajectory data management, graph data management and hybrid indexing structures.

### *3.1 Stream and trajectory data management*

Urban data often collected recurrently or even continuously (velocity) can lead to huge volumes of data collections that should be archived, organized (indexed) and maintained on persistence supports with efficient associated read and write mechanisms. Indexing and compression techniques are often applied to deal with data velocity and volume properties.

The continuous movement of an object is recorded in an approximate form as discrete samples of location points. A high sampling rate of location points generates accurate trajectories but will result in a massive amount of data leading to enormous overhead in data storage, communications, and processing. Thus, it is necessary to design data reduction techniques that compress the size of a trajectory while maintaining the utility of the trajectory. There are two major types of data reduction techniques running in batch after the data is collected (e.g., Douglas-Peucker algorithm [7]) or in an online mode as the data is being collected (such as the sliding window algorithm [12,16]). Trajectory reduction techniques are evaluated concerning three metrics: processing time, compression rate, and error measure (i.e., the deviation of an approximate trajectory from its original presentation). Recent research [18], has proposed solutions to the trajectory reduction through a hybrid spatial compression algorithm and error-bounded temporal compression algorithm. Chen et al. [5] propose to simplify a trajectory by considering both the shape skeleton and the semantic meanings of the trajectory [31,32]. For example, when exploring a trajectory (e.g., travel route) shared by a user, the places where she stayed, took photos, changed moving directions significantly would be more significant than other points. Consequently, points with an important semantic meaning should be given a higher weight when choosing representative points for a simplified trajectory.

*3.2 Graph Data Management*

Graphs are used to represent urban data, such as road networks, subway systems, social networks, and sensor networks. Graphs usually associated with a spatial property, resulting in many spatial graphs [36]. For example, the node of a road network has a spatial coordinate and each edge denoting a road segment has a spatial length. Graphs also contain temporal information, for instance, the traffic volume traversing a road segment changes over time, and the travel time between two landmarks is time-dependent: spatio-temporal graphs [36]. Queries like *"find the top-k tourist attractions around a user that are most popular in the past three months"*, can be asked on top of graphs.

*Hybrid Indexing Structures* are intended to well organize different data sources. For example, combining POIs, road networks, traffic, and human mobility data simultaneously. Hybrid structures can be used for indexing special regions. For instance, a city partitioned into grids by using a quad-tree-based spatial index (see Figure 2) where each leaf node (grid) of the spatial index maintains two lists storing the POIs and road segments. Then, each road segment ID points to two sorted lists: a list of taxi IDs sorted by their arrival time $ta$ at the road segment; a list of drop-off and pick-up points of passengers sorted by the pick-up time ($tp$) and drop-off time ($td$).

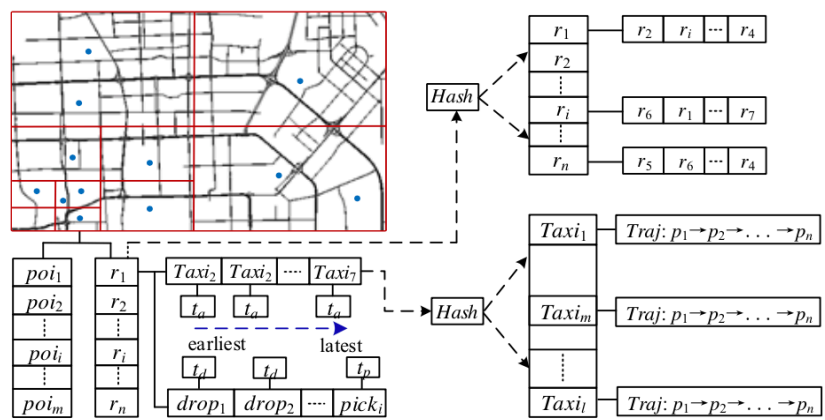

Figure 2 Hybrid index for organizing urban data [36]

Different kinds of index structures have been proposed to manage different types of data individually. Hybrid indexes that can simultaneously manage multiple types of data (e.g., spatial, temporal, and social media) for enabling efficient and effective learning of multiple heterogeneous data sources. In an urban computing scenario, it is usually needed to harness a variety of data and integrate them into a data-mining model. This calls for hybrid indexing structures that can well organize different data sources like hybrid indexing structure, which combines a spatial index, hash tables, sorted lists, and an adjacency list.

## 4. Querying Urban Data

Querying the actual location of a moving object has been studied extensively in moving object databases using 3DR-Tree [19] and MR-Tree [28]. Yet, sometimes queries must explore historical trajectories satisfying certain criteria, for example, retrieving the trajectories of tourists passing a given region and within a period. This corresponds to a spatiotemporal range query [23,24]), for example, taxi trajectories that pass a crossroad (i.e., a point query), or the trajectories that are similar to a query trajectory [6,20] (i.e., a trajectory query).

*Dealing with the uncertainty of a trajectory* refers to positioning moving objects while their locations can only be updated at discrete times. The location of a moving object between two updates uncertain because the time interval between two updates can exceed several minutes or hours. This can, however, save energy consumption and communication bandwidth.

*Map matching* is to infer the path that a moving object like a vehicle has traversed on a road network based on the sampled trajectory. Map-matching techniques dealing with high-sampling-rate trajectories have already been commercialized in personal navigation devices, while those for low-sampling-rate trajectories [15] are still considered challenging. According to Yuan et al. [30], given a trajectory with a sampling rate around 2 minutes per point, the highest accuracy of a map-matching algorithm is about 70%. When the time interval between consecutive sampling points becomes even longer existing map-matching algorithms do not work very well any longer [36]. Wei et al. [26] proposed to construct the most likely route passing a few sampled points based on many uncertain trajectories.

Krumm et al. and Xue et al. [13,29] propose solutions to predict a user's destination based on partial trajectories. More generally, a user's and other people's historical trajectories as well as other information, such as the land use of a location, can be used in destination prediction models.

Other important problems include observing a certain number of moving objects travelling a common sequence of locations in similar travel time where the locations in a travel sequence are not consecutive for finding *sequential patterns* from trajectories. Other approaches discover a group of objects that move together for a certain time period, under different patterns such as flock [8,9], convoy [10,11], swarm [14], traveling companion [21,22], and gathering [34,35,36,25]. These "group patterns", can be distinguished based on how the "group" is defined and whether they require the periods to be consecutive. For example, a flock is a group of objects that travel together within a disc of some user-specified size for at least k consecutive timestamps [10]. Li et al. [14] relaxed strict requirements on consecutive periods and proposed the pattern swarm, which is a cluster of objects lasting for at least k (possibly non-consecutive) timestamps.

## 5. Data and Knowledge Fusion

In urban computing scenarios, it is necessary to exploit a variety of heterogeneous data sources that need to be integrated. Then, it is necessary to fusion knowledge to explore and exploit datasets to extract insight and foresight of urban patterns and phenomena.

*Data fusion*. There are three major ways to achieve this goal:

- Fuse data sources at a feature level putting together the features extracted from different data sources into one feature vector. Beforehand and given the heterogeneity of data sources, a certain kind of normalization technique should be applied to this feature vector before feeding it into a data analytics model.
- Use different data at different stages. For instance, first partition an urban region, for example, a city into disjoint regions by major roads and then use human mobility data to glean the problematic configuration of a city's transportation network [33].
- Feed different datasets into different parts of a model simultaneously given a deep understanding of the data sources and algorithms applied to analyse them.

Building high-quality training datasets is one of the most difficult challenges of machine learning solutions in the real world. Disciplines like data mining, artificial intelligence and deep learning have contributed to building accurate models but, to do so, they require vastly larger volumes of training data. The traditional process for building a training dataset involves three tasks: data collection, data labelling and feature engineering. From the complexity standpoint, data collection is fundamentally trivial as most organizations understand what data sources they have. Feature engineering is getting to the point that is 70%-80% automated using algorithms. The real effort is in the data labelling stage. New solutions are emerging for combining strong and weak supervision methods to address data labelling.

*Knowledge fusion*. Data mining and machine-learning models dealing with a single data source have been well explored. However, the methodology that can learn mutually reinforced knowledge from multiple data sources is still missing. The fusion of knowledge does not mean simply putting together a collection of features extracted from different sources but also requires a deep understanding of each data source and effective usage of different data sources in different parts of a computing framework.

End-to-end urban computing scenarios call for the integration of algorithms of different domains. For instance, data management techniques with machine-learning algorithms must be combined to provide both efficient and effective knowledge discovery ability. Similarly, integrating spatio-temporal data management algorithms with optimization methods. Visualization techniques should be involved in a knowledge discovery process, working with machine-learning and data-mining algorithms.

## 6. Perspectives of the Role of Data Science for Making Urban Spaces Smart

This chapter discussed and described issues regarding data for enabling urban computing tasks that can lead to the design of smart urban territories. Having a data centred analysis of the problems and challenges introduced by urban computing exhibits the requirement to study data concerning different perspectives. First, the chapter characterised data produced within urban territories in terms of their mathematical properties (spatio-temporal), concerning the "semantics" of the entities composing urban territories (e.g., points of interest, roads, infrastructure) and also from the mobile entities that populate urban territories, like people, vehicles and the built environment. This variety of data are produced by producers with different characteristics, and approaches today use hardware, software and passive and active participation of people to generate phenomenological observations of urban territories. Finally, the chapter discusses how to create insight and foresight of important situations happening in urban territories, for example, computing trajectories of entities evolving in these territories observed in space and time; other social foresight of behaviours like popular POIs, the population of regions, etc.

The vision of urban computing—acquisition, integration, and analysis of big data to improve urban systems and life quality— is leading to smarter cities. Urban computing blurs the boundary between databases, machine learning, and visualization and even bridges the gap between different disciplines (e.g., computer sciences and civil engineering). To revolutionize urban sciences and progress, quite a few techniques still need to be explored such as the hybrid indexing structure for multimode data, the knowledge fusion across heterogeneous data sources, exploratory visualization for urban data, the integration of algorithms of different domains, and intervention-based analysis.